\RequirePackage{amsmath}
\documentclass[a4paper,epjc3]{svjour3}
\pdfoutput=1
\usepackage[utf8]{inputenc}
\usepackage{color,xcolor}
\definecolor{blue}{rgb}{0,0,0.6}
\usepackage[colorlinks,linkcolor=blue,urlcolor=blue,citecolor=blue]{hyperref}
\usepackage{amsmath,amssymb}
\usepackage{upgreek}
\usepackage{multirow}
\usepackage{tablefootnote}
\usepackage{cite}
\usepackage{graphicx}
\definecolor{maroon}{cmyk}{0, 0.87, 0.68, 0.32}
\definecolor{halfgray}{gray}{0.55}
\definecolor{ipython_frame}{RGB}{207, 207, 207}
\definecolor{ipython_bg}{RGB}{247, 247, 247}
\definecolor{ipython_red}{RGB}{186, 33, 33}
\definecolor{ipython_green}{RGB}{0, 128, 0}
\definecolor{ipython_cyan}{RGB}{64, 128, 128}
\definecolor{ipython_purple}{RGB}{170, 34, 255}

\usepackage{listings}
\lstdefinelanguage{iPython}{
    morekeywords={access,and,break,class,continue,def,del,elif,else,except,exec,finally,for,from,global,if,import,in,is,lambda,not,or,pass,print,raise,return,try,while},%
    morekeywords=[2]{abs,all,any,basestring,bin,bool,bytearray,callable,chr,classmethod,cmp,compile,complex,delattr,dict,dir,divmod,enumerate,eval,execfile,file,filter,float,format,frozenset,getattr,globals,hasattr,hash,help,hex,id,input,int,isinstance,issubclass,iter,len,list,locals,long,map,max,memoryview,min,next,object,oct,open,ord,pow,property,range,raw_input,reduce,reload,repr,reversed,round,set,setattr,slice,sorted,staticmethod,str,sum,super,tuple,type,unichr,unicode,vars,xrange,zip,apply,buffer,coerce,intern},%
    sensitive=true,%
    morecomment=[l]\#,%
    morestring=[b]',%
    morestring=[b]",%
    morestring=[s]{'''}{'''},
    morestring=[s]{"""}{"""},
    morestring=[s]{r'}{'},
    morestring=[s]{r"}{"},%
    morestring=[s]{r'''}{'''},%
    morestring=[s]{r"""}{"""},%
    morestring=[s]{u'}{'},
    morestring=[s]{u"}{"},%
    morestring=[s]{u'''}{'''},%
    morestring=[s]{u"""}{"""},%
    literate=
    {á}{{\'a}}1 {é}{{\'e}}1 {í}{{\'i}}1 {ó}{{\'o}}1 {ú}{{\'u}}1
    {Á}{{\'A}}1 {É}{{\'E}}1 {Í}{{\'I}}1 {Ó}{{\'O}}1 {Ú}{{\'U}}1
    {à}{{\`a}}1 {è}{{\`e}}1 {ì}{{\`i}}1 {ò}{{\`o}}1 {ù}{{\`u}}1
    {À}{{\`A}}1 {È}{{\'E}}1 {Ì}{{\`I}}1 {Ò}{{\`O}}1 {Ù}{{\`U}}1
    {ä}{{\"a}}1 {ë}{{\"e}}1 {ï}{{\"i}}1 {ö}{{\"o}}1 {ü}{{\"u}}1
    {Ä}{{\"A}}1 {Ë}{{\"E}}1 {Ï}{{\"I}}1 {Ö}{{\"O}}1 {Ü}{{\"U}}1
    {â}{{\^a}}1 {ê}{{\^e}}1 {î}{{\^i}}1 {ô}{{\^o}}1 {û}{{\^u}}1
    {Â}{{\^A}}1 {Ê}{{\^E}}1 {Î}{{\^I}}1 {Ô}{{\^O}}1 {Û}{{\^U}}1
    {œ}{{\oe}}1 {Œ}{{\OE}}1 {æ}{{\ae}}1 {Æ}{{\AE}}1 {ß}{{\ss}}1
    {ç}{{\c c}}1 {Ç}{{\c C}}1 {ø}{{\o}}1 {å}{{\r a}}1 {Å}{{\r A}}1
    {€}{{\EUR}}1 {£}{{\pounds}}1
    {^}{{{\color{ipython_purple}\^{}}}}1
    {=}{{{\color{ipython_purple}=}}}1
    {+}{{{\color{ipython_purple}+}}}1
    {*}{{{\color{ipython_purple}$^\ast$}}}1
    {/}{{{\color{ipython_purple}/}}}1
    {+=}{{{+=}}}1
    {-=}{{{-=}}}1
    {*=}{{{$^\ast$=}}}1
    {/=}{{{/=}}}1,
    literate=
    *{-}{{{\color{ipython_purple}-}}}1
     {?}{{{\color{ipython_purple}?}}}1,
    identifierstyle=\color{black}\ttfamily,
    commentstyle=\color{ipython_cyan}\ttfamily,
    stringstyle=\color{ipython_red}\ttfamily,
    keepspaces=true,
    showspaces=false,
    showstringspaces=false,
    rulecolor=\color{ipython_frame},
    frame=single,
    frameround={t}{t}{t}{t},
    framexleftmargin=6mm,
    numbers=left,
    numberstyle=\tiny\color{halfgray},
    backgroundcolor=\color{ipython_bg},
    basicstyle=\footnotesize\ttfamily,
    keywordstyle=\color{ipython_green}\ttfamily,
    aboveskip=1.2em,
    belowskip=1.2em,
}

\usepackage{bold-extra}

\newcommand{\flavio}{\texttt{flavio}}
\newcommand{\wilson}{\texttt{wilson}}

\begin{document}

\title{\raisebox{-2.2mm}{\includegraphics[width=0.9cm]{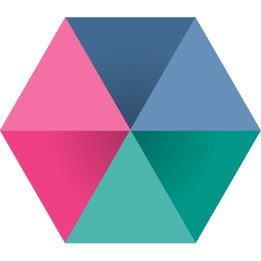}}~ \huge \color{black!75}\texttt{flavio}}
\subtitle{a Python package for flavour and precision phenomenology in the Standard Model and beyond}
\author{David~M.~Straub}
\institute{Excellence Cluster Universe, TUM, Boltzmannstr.~2,
85748~Garching, Germany}
\date{}
\maketitle

\begin{abstract}
\flavio{} is an open source tool for phenomenological analyses in flavour physics and other precision observables in the Standard Model and beyond.
It consists of a library to compute predictions for a plethora of
observables in quark and lepton flavour physics
and electroweak precision tests,
a database of experimental measurements of these observables,
a statistics package that allows to construct Bayesian and frequentist likelihoods, and of convenient plotting and visualization routines.
New physics effects are parameterised as Wilson coefficients of dimension-six operators in the weak effective theory below
the electroweak scale or the Standard Model EFT above it.
At present, observables implemented include numerous rare $B$ decays (including angular observables of
exclusive decays, lepton flavour and lepton universality violating $B$ decays),
meson-antimeson mixing observables in the $B_{d,s}$, $K$, and $D$ systems,
tree-level semi-leptonic $B$, $K$, and $D$ decays (including possible lepton universality violation),
rare $K$ decays,
lepton flavour violating $\tau$ and $\mu$ decays,
$Z$ pole electroweak precision observables,
the neutron electric dipole moment,
and anomalous magnetic moments of leptons.
Not only central values but also theory uncertainties of all observables can be computed.
Input parameters and their uncertainties can be easily modified by the user.
Written in Python, the code does not require compilation and can be
run in an interactive session.
This document gives an overview of the features as of version 1.0 but does not represent a manual.
The full documentation of the code can be found in its web site.
\end{abstract}

\newpage
\tableofcontents

\section{Introduction}

The absence of any new particles beyond the Standard Model (SM) in high-energy
collisions at the LHC highlights the need to probe the SM in low-energy experiments,
at the high-intensity frontier, where new physics (NP) might leave its imprint
through non-standard interactions among known particles, corresponding
to higher-dimensional operators in an effective field theory (EFT).
In this context, flavour physics plays a pivotal role for several reasons.
First, flavour-changing neutral current (FCNC) processes such as rare meson
decays or meson-antimeson mixing are forbidden at tree-level in the SM, so they
are particularly sensitive to NP effects. Second, flavour physics is required
to determine most of the free parameters in the SM, notably
the three angles and the phase of the Cabibbo-Kobayashi-Maskawa (CKM) matrix.


A major challenge in flavour physics is the disentanglement of non-perturbative
strong-interaction effects, affecting many hadronic low-energy observables,
from possible NP contributions.
In recent years, several apparent deviations from the SM have been observed
in flavour physics, e.g.\
in angular observables of $B\to K^*\mu^+\mu^-$ \cite{Aaij:2015oid} or lepton flavour
universality ratios in $B\to K^{(*)}\ell^+\ell^-$ \cite{Aaij:2014ora,Aaij:2017vbb} and $B\to D^{(*)}\ell\nu$ \cite{Lees:2013uzd}.
In all of these cases, it is important to scrutinize the SM predictions and
uncertainties, but also to investigate possible NP explanations.
However, often the complicated dependence of flavour physics observables on the
Wilson coefficients is a major obstacle in connecting flavour phenomenology and
model building. In addition, it is difficult comparing different
theoretical approaches with different methods, parameter choices, and
statistical frameworks.
This highlights the virtue of open source codes that allow
to compute predictions for observables in the flavour sector
in terms of NP effects parameterised by Wilson coefficients
of dimension-6 operators,
properly taking into account all hadronic and other uncertainties.

\flavio{} is an open source Python package that contains
\begin{itemize}
  \item a vast library of observables in flavour physics, electroweak precision tests, and other low energy observables, as function of dimension-6 Wilson coefficients in an EFT above or below the electroweak scale,
  \item a database of experimental measurements of these observables,
  \item a module to automatically construct likelihoods in terms of parameters and Wilson coefficients using these measurements,
  \item various plotting and visualization routines.
\end{itemize}
\flavio{} overlaps in scope with  several other  open source  packages  in HEP, most notably
\begin{itemize}
  \item \texttt{EOS} \cite{EOS}, written in C++, that allows to compute a large number of observables in $B$ physics and contains Bayesian sampling routines,
  \item \texttt{HEPfit} \cite{HEPfit}, written in C++, that contains a library not only of flavour physics but also electroweak precision tests and Higgs physics, implements specific dynamical NP models, and is focused on performing Bayesian analyses,
  \item \texttt{SuperIso} \cite{Mahmoudi:2009zz}, written in C, that allows to  compute many observables in $B$ physics in a general EFT or specific NP models.
\end{itemize}
The focus of \flavio{} is different from these packages in some respects. In particular,
\begin{itemize}
  \item it is written entirely in Python, allowing for a relatively easy extension, modification of the code at runtime,
  and interactive execution,
  \item thanks to Python it is trivial to install, not requiring compilation,
  \item it aims to include as many observables as possible where NP effects can be expressed in terms of dimension-6 Wilson coefficients, not necessarily limited to $B$ (or flavour) physics,
  \item it focuses on NP effects purely within an EFT rather than implementing specific NP models, leaving this to dedicated programmes (see e.g. \cite{Porod:2014xia, Evans:2016lzo}),
  \item it attempts to be general when it comes to the statistical approach, not committing to a Bayesian or frequentist framework.
\end{itemize}
Clearly, all the existing packages have advantages of their own
and the crosscheck between them has proven to be very useful.

First released in early 2016, \flavio{} has already used in a number of publications by various groups
\cite{Paul:2016urs, Bhattacharya:2016mcc, Altmannshofer:2017wqy, Altmannshofer:2017fio, Alok:2017jgr, Altmannshofer:2017yso, DAmico:2017mtc, Sala:2017ihs, DiChiara:2017cjq, Alok:2017jaf, Alok:2017sui, Poh:2017tfo, Datta:2017ezo, Medina:2017bke, Blake:2017fyh, Albrecht:2017odf, Altmannshofer:2017bsz, Fuyuto:2017sys, Sannino:2017utc, Jung:2018lfu, Kumar:2018kmr, Sala:2018ukk}.
This note describes \flavio{} as of version 1.0.
The code is in active development, which takes place in a public
GitHub repository \cite{GitHub}.
Code contributions from the community (via \textit{pull requests}) are
encouraged.
The full documentation can be found on the \flavio{}
website \cite{website}.
Tables~\ref{tab:obs} and \ref{tab:obs2} list the processes and observables implemented so far
along with the references that were most relevant for the
implementation of the process\footnote{Note that this is not necessarily the
reference to the first discussion of a particular observable in the literature
but in some cases a review article. For original references, please consult
reviews of the field, e.g.\ \cite{Buras:2011we,Blake:2016olu}.}.

\begin{table}[tbp]
\small
\renewcommand{\arraystretch}{1.2}
\centering
\begin{tabular}{llcc}
\hline
Category & Process & Observables & References \\
\hline
\multirow{3}{*}{Meson mixing}
& $B^0\leftrightarrow \bar B^0$ & $\Delta M_d$, $a_\text{fs}^d$ &
\cite{Nierste:2009wg,Artuso:2015swg,Bazavov:2016nty}
\\
&$B_s\leftrightarrow \bar B_s$ & $\Delta M_s$, $a_\text{fs}^s$ &
\cite{Nierste:2009wg,Artuso:2015swg,Bazavov:2016nty}\\
&$K^0\leftrightarrow \bar K^0$ & $|\epsilon_K|$ &
\cite{Brod:2011ty} \\
&$D^0\leftrightarrow \bar D^0$ & $x$, $y$, $\phi$, $q/p$,&
 \\
&& $x_{12}$, $y_{12}$, $\phi_{12}$, $x_{12}^\text{Im}$ &
\cite{Amhis:2016xyh} \\
\hline
\multirow{2}{*}{Non-lept. $B$ decays} &$B^0\to\psi K_S$ & $S_{\psi K_S}$ \\
&$B^0\to\psi \phi$ & $S_{\psi \phi}$ \\
\hline
\multirow{4}{*}{Radiative $B$ decays}
&$B\to X_{s,d}\gamma$ & BR, $A_\text{CP}$ &
\cite{Gambino:2000fz,Misiak:2015xwa,Czakon:2015exa}\\
&$B^+\to K^{*+}\gamma$ & BR, $A_\text{CP}$ &
\cite{Beneke:2001at,Straub:2015ica}\\
&$B^0\to K^{*0}\gamma$ & BR, $S_{K^*\gamma}$, $A_\text{CP}$ &
\cite{Beneke:2001at,Ball:2006eu,Straub:2015ica}\\
&$B_s\to \phi\gamma$ & BR, $S_{\phi\gamma}$, $A_{\Delta\Gamma}$ &
\cite{Beneke:2001at,Muheim:2008vu,Straub:2015ica}\\
\hline
Rare lept. $B$ decays
&$B_{s,d}\to \ell^+\ell^-$ & BR, $A_{\Delta\Gamma}$ &
\cite{DeBruyn:2012wk,Bobeth:2013uxa,Becirevic:2016zri}\\
\hline
\multirow{6}{*}{Rare SL $B$ decays}
&$B^+\to \ell^+\nu$ & BR &
\\
&$B^{+,0}\to K^{*+,0}\nu\bar\nu$ & BR, $F_L$  &
\cite{Altmannshofer:2009ma,Brod:2010hi,Straub:2015ica,Buras:2014fpa}\\
&$B^{+,0}\to K^{+,0}\nu\bar\nu$ & BR  &
\cite{Altmannshofer:2009ma,Brod:2010hi,Straub:2015ica,Buras:2014fpa}\\
&$B^{+,0}\to \rho^{+,0}\nu\bar\nu$ & BR  &
\cite{Altmannshofer:2009ma,Brod:2010hi,Straub:2015ica}\\
&$B^{+,0}\to \pi^{+,0}\nu\bar\nu$ & BR  &
\cite{Brod:2010hi,Bailey:2014fpx,Bailey:2015nbd}\\
&$B^{+,0}\to K^{*+,0}\ell^+\ell^-$ & BR, $S_i$, $A_i$, $P_i^{(\prime)}$ &
\cite{Beneke:2001at,Beneke:2004dp,Greub:2008cy,Altmannshofer:2008dz,Descotes-Genon:2013vna,Altmannshofer:2014rta,Horgan:2015vla,Gratrex:2015hna,Straub:2015ica}\\
&$B^{+,0}\to K^{+,0}\ell^+\ell^-$ & BR, $F_H$, $A_\text{FB}$ &
\cite{Greub:2008cy,Bartsch:2009qp,Altmannshofer:2014rta,Gratrex:2015hna,Bailey:2015dka}\\
&$B_s\to \phi\ell^+\ell^-$ & BR, $F_L$, $S_i$ &
\cite{Horgan:2015vla,Descotes-Genon:2015hea,Straub:2015ica} \\
&$\Lambda_b\to \Lambda\ell^+\ell^-$ & BR, $F_L$, $A_\text{FB}^{\ell,h,\ell h}$ &
\cite{Detmold:2016pkz, Meinel:2016grj} \\
\hline
\multirow{4}{*}{SL tree-level $B$ dec.}
&$B^{0,+}\to\pi^{+,0}\ell\nu$ & BR &
\cite{Gratrex:2015hna,Lattice:2015tia}\\
&$B^{0,+}\to\rho^{+,0}\ell\nu$ & BR &
\cite{Gratrex:2015hna,Straub:2015ica}\\
&$B^{0,+}\to D^{*+,0}\ell\nu$ & BR, $\text{BR}_{L,T}$,  $F_L$, $d\text{BR}/d\chi_i$ &
\cite{Gratrex:2015hna,Lattice:2015rga,Dungel:2010uk,Abdesselam:2017kjf}\\
&$B^{0,+}\to D^{+,0}\ell\nu$ & BR &
\cite{Gratrex:2015hna,Lattice:2015rga}\\
&$B^{-}\to\omega\ell^-\bar\nu$ & BR &
\cite{Gratrex:2015hna,Straub:2015ica}\\
&$B_s\to K^{*+}\ell^-\bar\nu$ & BR &
\cite{Gratrex:2015hna,Straub:2015ica} \\
\hline
\multirow{2}{*}{Lep. tree-level $B$ dec.}
&$B^+\to\ell\nu$ & BR \\
&$B_c\to\ell\nu$ & BR \\
\hline
Rare $K$ decays &$K^{+,}{}_L\to \pi^{+,0}\nu\bar\nu$ & BR  &
\cite{Buras:2005gr,Brod:2010hi}\\
\hline
Non-leptonic $K$ decay &$K\to \pi\pi$ & $\varepsilon'/\varepsilon$  &
\cite{Buras:2015yba,Aebischer:2018rrz,Aebischer:2018quc,Aebischer:2018csl,Blum:2015ywa,Bai:2015nea}\\
\hline
\multirow{2}{*}{SL tree-level $K$ dec.}
&$K^+\to\ell^+\nu$ & BR &
\cite{Cirigliano:2007ga,Gonzalez-Alonso:2016etj}\\
&$K^{+,}{}_L\to\pi^{0,+}\ell^-\bar\nu$ & BR &
\cite{Bernard:2009zm,Antonelli:2010yf,Baum:2011rm,Gonzalez-Alonso:2016etj}\\
\hline
$\pi$ decays
&$\pi^+\to e^+\nu$ & BR &
\cite{Cirigliano:2007ga,Gonzalez-Alonso:2016etj}\\
\end{tabular}
\caption{Quark flavour physics processes and observables implemented in \flavio{} 1.0.
Abbreviations: SL -- semi-leptonic, BR -- branching ratio.
In all processes involving final-state leptons, arbitrary
violation of lepton flavour universality or charged lepton flavour is possible.}
\label{tab:obs}
\end{table}

\begin{table}[tbp]
\small
\renewcommand{\arraystretch}{1.2}
\centering
\begin{tabular}{llcc}
\hline
Category & Process & Observables & References \\
\hline
\multirow{4}{*}{LFV $\tau$ decays}
&$\tau\to\ell\gamma$ &  BR & \cite{Brignole:2004ah}\\
&$\tau\to 3\mu$ &  BR &\cite{Brignole:2004ah}\\
&$\tau\to \mu ee$ &  BR &\cite{Brignole:2004ah}\\
&$\tau\to \rho\ell$ &  BR &\cite{smelli}\\
&$\tau\to \phi\ell$ &  BR &\cite{smelli}\\
\hline
\multirow{4}{*}{Tree-level $\tau$ decays}
&$\tau\to\ell\nu\nu$ &  BR & \cite{Pich:2013lsa}\\
&$\tau\to K\nu$ &  BR &\\
&$\tau\to \pi\nu$ &  BR &\\
\hline
\multirow{2}{*}{LFV $\mu$ decays}
&$\mu\to e\gamma$ &  BR &\cite{Brignole:2004ah}\\
&$\mu\to 3e$ &  BR &\cite{Brignole:2004ah}\\
\hline
\multirow{2}{*}{$Z$ prod. \& decay}
&$Z\to f\bar f$ &  $\Gamma_Z, \Gamma_f,  A, A_\text{FB}, R_f$ &\cite{Efrati:2015eaa, Brivio:2017vri}\\
&$e^+e^-\to Z\to q\bar q$ &  $\sigma_\text{had}$ &\cite{Efrati:2015eaa, Brivio:2017vri}\\
\hline
\multirow{2}{*}{$W$ prod. \& decay}
&$W\to \ell\nu$ &  BR &\cite{Efrati:2015eaa, Brivio:2017vri}\\
&$W\to ff'$ & $\Gamma_W$ &\cite{Efrati:2015eaa, Brivio:2017vri}\\
&$m_W$& &\cite{Brivio:2017vri}\\
\hline
EDMs
&$d_n$ &  &
\cite{Demir:2002gg,Fuyuto:2013gla,Bhattacharya:2015wna}\\
\hline
MDMs
&$a_e, a_\mu, a_\tau$ &&
\cite{Eidelman:2007sb,Davier:2017zfy,Jegerlehner:2018zrj}
\\
\hline
other
&$\nu_\mu N\to \nu_\ell\mu^+\mu^-N$ &
$R_\text{trident}$ &
\cite{Altmannshofer:2014pba}
\\
\hline
\end{tabular}
\caption{Processes and observables beyond quark flavour physics
implemented in \flavio{} 1.0.
Abbreviations: BR -- branching ratio, LFV -- lepton flavour violation,
EDM -- electric dipole moment, MDM -- (anomalous) magnetic dipole moment.}
\label{tab:obs2}
\end{table}

\section{Getting started}

\subsection{Installation}

\flavio{} requires Python version 3.5 or above and runs on Linux, Mac, and Windows.
Being hosted at the Python Package Index \cite{PyPI},
it can be installed with a single command using the Python
package manager \texttt{pip}:\footnote{Depending on the system, the Python 3 executable might have a name different from \texttt{python3}.}
\begin{lstlisting}[language=iPython]
$ python3 -m pip install flavio --user
\end{lstlisting}
Using the sampling and plotting functionality requires additional packages that can be installed with
\begin{lstlisting}[language=iPython]
$ python3 -m pip install flavio[sampling,plotting] --user
\end{lstlisting}
When a new version is released, the package can be easily upgraded by the command
\begin{lstlisting}[language=iPython]
$ python3 -m pip install flavio --upgrade
\end{lstlisting}
Please report any installation problems on the Github issue page\footnote{\url{https://github.com/flav-io/flavio/issues}}.

\subsection{Using flavio}

As for any Python package, there are several ways of using \flavio{}:
\begin{itemize}
 \item As library imported in another script,
 \item In a simple interactive command-line interface (obtained by calling \texttt{python3} in a terminal),
 \item In a more sophisticated and user-friendly interactive command-line interface (IPython \cite{IPythonweb}),
 \item In an interactive notebook interface, similar to Mathematica notebooks,
 running in a browser (Jupyter Notebook \cite{Jupyterweb}).
\end{itemize}
For the installation of IPython or Jupyter, please consult the respective documentations.

When using \flavio{} in a parameter scan or similar batch
operation, it is advisable to implement the loop in Python
and to import the package only once, as the loading of the
experimental measurements etc.\ leads to a significant, one-time overhead on import.

\section{Overview of features}

\subsection{SM predictions and uncertainties}

The most basic functionality of \flavio{} is to compute predictions for flavour
observables in the Standard Model (SM) and beyond.
To get started, just import the package:
\begin{lstlisting}[language=iPython]
import flavio
\end{lstlisting}
Observables are identified by strings. For instance,
the time-integrated branching ratio of $B_s\to\mu^+\mu^-$
is identified by the string \texttt{'BR(Bs->mumu)'}. To get the central
value of its SM prediction, use
\begin{lstlisting}[language=iPython]
flavio.sm_prediction('BR(Bs->mumu)')
\end{lstlisting}
The complete list of observables and their string names is
available in the online documentation \cite{website}.

Some observables depend on additional parameters, like a dilepton invariant
mass squared $q^2$. To get the differential branching ratio of $B^+\to K^+\mu^+\mu^-$
at $q^2=3\,\text{GeV}^2$, use
\begin{lstlisting}[language=iPython]
flavio.sm_prediction('dBR/dq2(B+->Kmumu)', q2=3)
\end{lstlisting}
Note that all dimensionful quantities in \flavio{} are always assumed to be in units
of GeV to the appropriate power.

To get the uncertainty of the SM prediction of the $B_s\to\mu^+\mu^-$ branching
ratio, use
\begin{lstlisting}[language=iPython]
flavio.sm_uncertainty('BR(Bs->mumu)')
\end{lstlisting}
Uncertainties are computed by drawing random values
for all input parameters according to their probability distribution, computing the
observable for all of them, and extracting the standard deviation of the spread
of values. Since this procedure involves random numbers, its precision depends
on the number of random numbers. By default, the function uses 100 iterations.
This can be increased by
specifying the optional parameter \texttt{N=200}, for instance. Note that the relative
error of the uncertainty as a function of $N$ is given by
$\frac{\Delta \sigma}{\sigma} = 1/\sqrt{2N}$.

\subsection{Predictions in the presence of new physics}

\flavio{} does not implement any specific NP models. Rather, it allows
the user to specify NP contributions in the form of Wilson coefficients of dimension-6
operators, either in the weak effective field theory (WET) below the electroweak scale or in the Standard Model effective field
theory (SMEFT) above it. Since version 0.28,
the renormalization group evolution, matching, and basis translation
is performed using the \wilson{} package \cite{Aebischer:2018bkb}
building on the Wilson coefficient exchange format (WCxf) \cite{Aebischer:2017ugx}.
The default WCxf basis used by \flavio{} for SMEFT is \texttt{Warsaw},
for WET  it is \texttt{flavio}. Other bases can be used
if a corresponding translator is defined in the \texttt{wcxf}
Python package.

Using the \wilson{} package, a new ``parameter point'' in the space of the EFT maps to an instance
of the class \texttt{wilson.Wilson}\footnote{%
The class \texttt{flavio.WilsonCoefficients} is a subclass of
\texttt{wilson.Wilson} defining additional methods for backward
compatibility.
}.
By default, all Wilson coefficients vanish. Note that these Wilson coefficients
are defined to be \textit{the new physics contributions only}, so if they vanish
it means we are in the Standard Model.
To go beyond the SM, the user must specify the values of the Wilson
coefficients at some scale. For instance, to get a universal NP contribution to
$B^0$ and $B_s$ mixing
via the operators $(\bar q_R\gamma^\mu b_R)^2$ with $q=d,s$,
\begin{lstlisting}[language=iPython]
from wilson import Wilson
w = Wilson({'CVRR_bdbd': 1e-10, 'CVRR_bsbs': 1e-10},
           scale=160, eft='WET', basis='flavio')
\end{lstlisting}
The Wilson coefficients are referred to by their names. A
complete list of operators and Wilson coefficients is available on the WCxf web site \cite{WCxfweb} or in the \flavio{} documentation.

Having defined a NP scenario in this way, the central
value of any observable can now be obtained by calling the function
\texttt{np\char`_prediction}, e.g.
\begin{lstlisting}[language=iPython]
flavio.np_prediction('DeltaM_s', w)
\end{lstlisting}

\subsection{Experimental measurements}

\flavio{} contains a database of experimental measurements
that allows to compare the computed predictions to the data
and to construct sophisticated likelihoods for parameter
inference (see section~\ref{sec:llh}). The measurements are defined in a central YAML
file and new measurements can be easily added by the user.
Internally, measurements are defined as (univariate or multivariate) probability distributions in the space of
observables. To make this possible, the \texttt{flavio.statistics}
submodule defines a number of one- and multidimensional
probability distributions. This allows in particular
to deal with non-Gaussian and correlated measurements.

A simple use of the experimental measurements is to return the probability distribution of a single observable (combining multiple measurements if applicable)
\begin{lstlisting}[language=iPython]
d = flavio.combine_measurements('<Rmue>(B0->K*ll)',
                                q2min=1.1, q2max=6.0)
\end{lstlisting}
The properties of this distribution can bow be studied by methods such as \texttt{d.central\char`_value}, \texttt{d.ppf} (percent point function, i.e.\ the inverse cumulative distribution function), \texttt{d.error\char`_right}, etc. The PDF can also be plotted using the \texttt{flavio.plots} submodule:
\\\noindent
\begin{minipage}{\linewidth}
\begin{lstlisting}[language=iPython]
import flavio.plots as fpl
fpl.pdf_plot(d)
fpl.plt.xlabel('$R_{K^*}$')
\end{lstlisting}
\end{minipage}
\begin{center}
\includegraphics[width=0.5\textwidth]{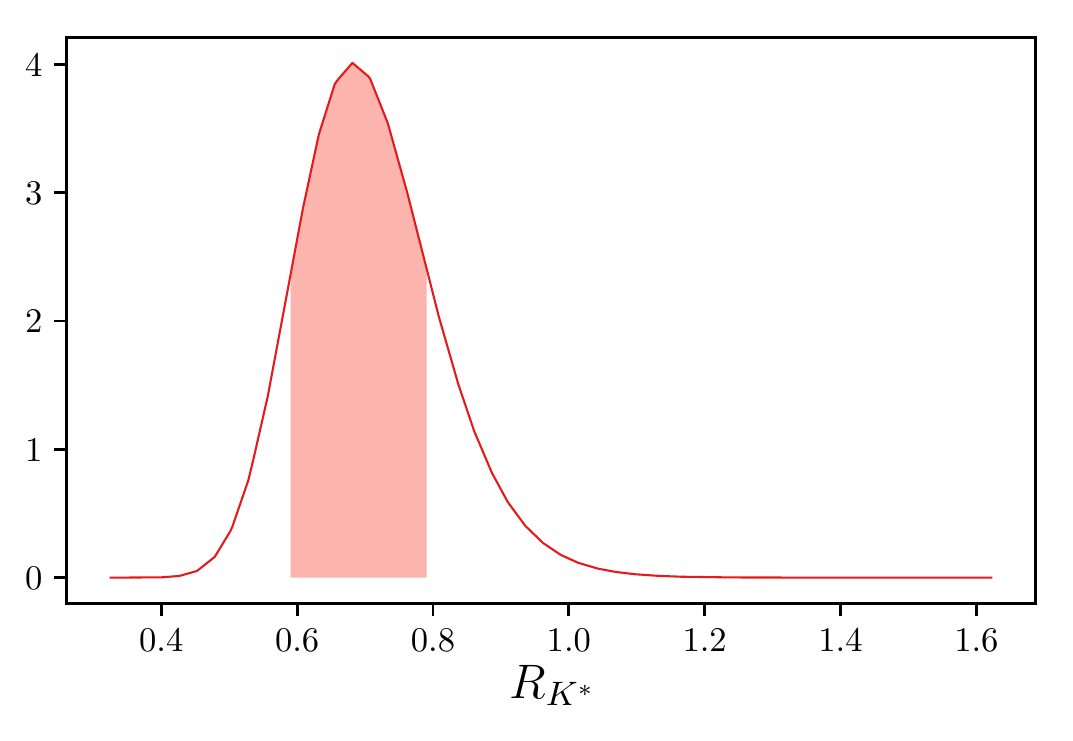}
\end{center}

For binned measurements of observables depending on a kinematic variable (e.g.\ $q^2$), there are dedicated functions to plot all existing measurements and also the theory predictions. For instance, to compare the SM prediction to an experimental measurement of the $B\to D\ell\nu$ differential branching ratio, one can use\footnote{%
The \texttt{divide\char`_binwidth} option is necessary
to correctly compare the (dimensional) partial branching ratios
to the (dimensionful) $q^2$-differential branching ratio.}
\begin{lstlisting}[language=iPython]
fpl.bin_plot_exp('<BR>(B+->Dlnu)', divide_binwidth=True)
fpl.bin_plot_exp('<BR>(B+->Denu)', divide_binwidth=True)
fpl.diff_plot_th('dBR/dq2(B+->Denu)', 0.01, 11.6, label='SM')
fpl.plt.legend()
\end{lstlisting}
\begin{center}
\includegraphics[width=0.5\textwidth]{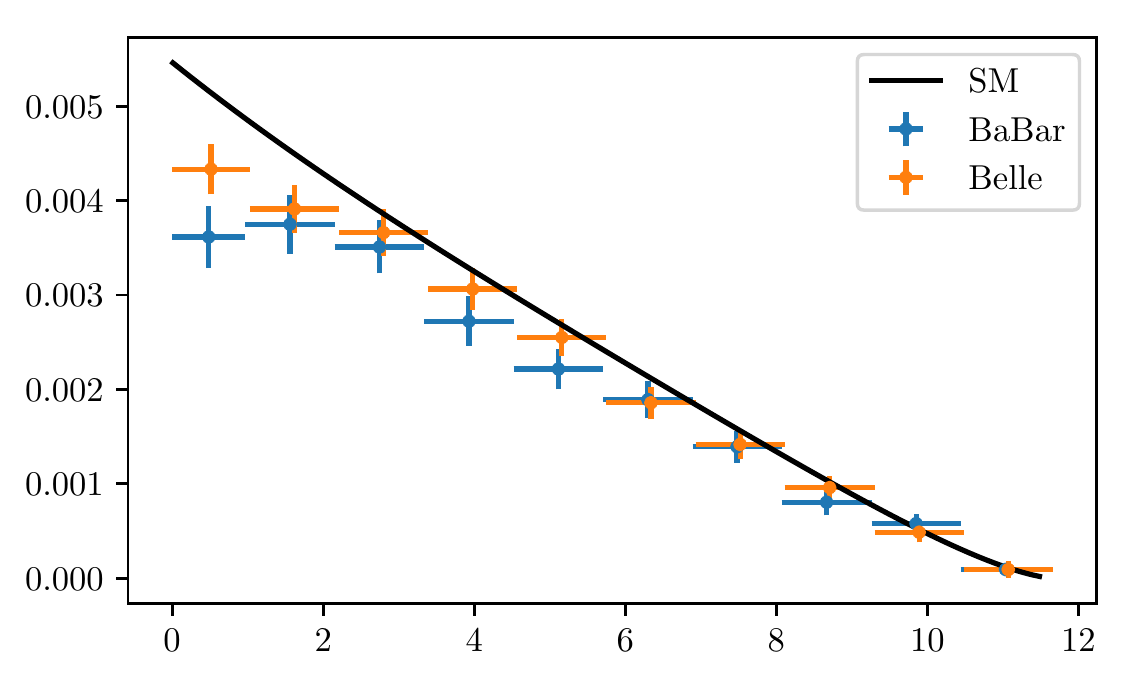}
\end{center}
Labels etc.\ can be added using the usual functions of the \texttt{matplotlib} library.

\section{Constructing likelihoods}\label{sec:llh}

Since \flavio{} contains a submodule with statistical routines
and a database of experimental measurements, it is straightforward
to construct likelihoods that can be used for Bayesian or
frequentist inference and for testing NP models.

\subsection{General likelihoods}

The basic class for general likelihoods is the \texttt{Likelihood}
class in the \texttt{flavio.\allowbreak{}statistics.\allowbreak{}likelihood} module.
It provides access to a likelihood function defined as
\begin{equation}
L(\vec C, \vec{\theta})
= \prod_i L_\text{exp}^i\left(\vec{O}^\text{exp}_{i}, \vec{O}^\text{th}_i\left(\vec C, \vec{\theta}\right)\right)
\times L_\theta(\vec{\theta})\,,
\label{eq:llh}
\end{equation}
where
\begin{itemize}
  \item $\vec C$ are a set of WET or SMEFT Wilson coefficients,
  \item $\vec{\theta}$ a set of parameters,
  \item $\vec{O}^\text{th}$ the theory predictions for the observables,
  \item $\vec{O}^\text{exp}$ the experimental measurements of the observables,
  \item $L_\text{exp}^i$ the probability distributions that are defined in the database of experimental measurements,
  \item $L_\theta(\vec{\theta})$ the probability distribution associated to a parameter that is defined in an instance of the \texttt{flavio.ParameterConstraints} class.
\end{itemize}
The simplest possiblity is to include in $L_\theta(\vec{\theta})$
all the constraints present by default in \flavio{} and
to include all the known measurements in $L_\text{exp}^i$.
Then, a likelihood instance can be simply defined by specifying
the observables to be included, e.g.
\begin{lstlisting}[language=iPython]
from flavio.statistics.likelihood import Likelihood
llh = Likelihood(observables=['DeltaM_d', 'DeltaM_s'])
\end{lstlisting}
The logarithm of the likelihood $\ln L$ can now be evaluated by calling the \texttt{log\char`_likelihood} method of the
instance, specifying a dictionary of parameter values and
the Wilson coefficients as an instance of \texttt{wilson.Wilson}.
More options are described in the online documentation.
A particularly useful feature is to load likelihood definitions from a YAML file.

At this point, a word of caution is in order: using the default option to include all constraints on parameters and all existing measurements can lead to inconsistent results in several cases, e.g.
\begin{itemize}
  \item including constraints on parameters that come from measurements included in the likelihood (e.g. the default constraint on $V_{ub}$ that comes from $B\to\pi\ell\nu$),
  \item including multiple measurements that are not independent of each other.
\end{itemize}
It is the responsibility of the user to ensure a consistent treatment.

\subsection{Fast likelihoods}\label{sec:flh}

While the likelihood \eqref{eq:llh} is very general and can be used for Bayesian or frequentist analyses,
it has the drawback that it depends on a large number of
parameters that must be varied to correctly account for
theoretical uncertainties, but might not be of prime interest,
i.e., nuisance parameters.
Since marginalizing or profiling over nuisance parameters
can be computationally demanding, \flavio{} also provides an
alternative way of constructing a likelihood, the \texttt{FastLikelihood}.
It is based on the approximation of assuming the likelihood to be of the form
\begin{gather}
-2\ln L(\vec C)
= \sum_i\vec{x}_i^T(\vec C)
\left[C_\text{exp} + C_\text{th}(\vec\theta_0)\right]^{-1}
\vec{x}_i(\vec C) \,,
\\
\vec{x}_i(\vec C) = \vec{O}^\text{exp}_i - \vec{O}^\text{th}_i(\vec C, \vec\theta_0) \,.
\end{gather}
where $C_\text{exp}$ is a covariance matrix of experimental measurements and
$C_\text{th}$ a covariance matrix of theory predictions in the SM for the central values ($\vec\theta_0$) of the theory parameters.
$C_\text{th}$ is obtained from randomly sampling the observables for theory parameters distributed according to their PDFs, while $C_\text{exp}$ is obtained from approximating the true experimental PDFs as (multivariate) Gaussians.

This approach has the main advantage that it yields a likelihood independent of nuisance parameters, but the time-consuming step (namely evaluating the theoretical covariance $C_\text{th}$)
is independent of the data. In particular, this makes this approach very powerful for fast inference after a change in experimental data.

However, its validity relies on a number of assumptions,
\begin{itemize}
\item the experimental uncertainties are approximated as Gaussian,
\item the theory uncertainties (at the level of observables) are approximated as Gaussian,
\item the covariances are assumed to be weakly dependent on $\vec\theta$.
\end{itemize}
The last point is by far the strongest assumption and its validy has to be checked whenever the method is employed.

This method was employed in several model-independent analyses of new physics in $b\to s\ell\ell$ transitions \cite{Altmannshofer:2014rta,Altmannshofer:2017fio,Altmannshofer:2017yso}.

\subsection{Fits}

While the fast likelihood discussed in \ref{sec:flh} allows to directly investigate the likelihood function in the space of
NP Wilson coefficients without having to deal with nuisance
parameters, a more proper statistical treatment of nuisance
parameters is desirable and even mandatory in some cases, e.g.\
when one of the assumptions underlying the fast likelihood is
not satisfied. One can then choose either a Bayesian or
a frequentist statistical framework to deal with the nuisance
parameters. \flavio's likelihood module is written in a
general way such that both approaches are supported

In the Bayesian approach, $L_\theta$ in \eqref{eq:llh}
would be interpreted as the prior probability distribution
of the parameters and the function $L$ on the left-hand side
would be the posterior probability distribution, up to normalization.
To obtain a distribution only for a subset of the parameters, the others have to be marginalized, i.e. integrated over.
This can be done efficiently in high dimensions using Monte Carlo methods, e.g. using nested sampling or Markov chains (MCMC).

In the frequentist approach, the parameters do not have prior probabilities associated to them; however, they can be subject to external, direct measurements that enter the likelihood separately. Even constraints that are purely
theoretical can be formally treated in the same way\footnote{%
See \cite{Charles:2016qtt} for a general discussion of the theory uncertainties in the frequentist approach.
}. Then, the factor $L_\theta$ in \eqref{eq:llh} simply
corresponds to this likelihood of pseudo-measurements of theory
parameters.
In \flavio{}, all nuisance parameters have a probability density function associated to them and the form of this PDF can be chosen to mimick different treatments of theory uncertainties. For instance, a normal distribution would correspond to treating the theory uncertainties like a statistical uncertainty, while a uniform distribution would lead to a treatment similar to the Rfit \cite{Hocker:2001xe} scheme.

In \flavio{}, marginalization and profiling routines are contained in the
\texttt{flavio.\allowbreak{}statistics.\allowbreak{}fitters}
submodule. Currently, it contains
\begin{itemize}
  \item a wrapper around the \texttt{emcee} \cite{ForemanMackey:2012ig} MCMC sampler,
  \item a wrapper around the \texttt{pypmc} \cite{pypmc} MCMC sampler,
  \item a custom frequentist likelihood profiler added in v0.22.
\end{itemize}
These routines have already been used in published papers (see e.g.\ \cite{Paul:2016urs} for an application of a Bayesian MCMC
and \cite{Jung:2018lfu} for an application of the likelihood profiler).
However due to the new \texttt{likelihood} module introduced in version 1.0, the interface is subject to change, which is why the reader is referred to the online documentation for updated usage instructions.

\section{Further information and outlook}

The purpose of this document was to give an overview of the features of the \flavio{} package at a specific point in time corresponding to the release of version 1.0. A detailed and updated documentation of the evolving package can be found on the \flavio{} web site~\cite{website}. It also contains the link to an interactive tutorial running as a Jupyter notebook on a virtual machine (powered by Binder) that allows the user to try the package in a browser without installing anything.

Since \flavio{} is open source, code contributions and improvements are highly welcome. In particular, many precision observables are still missing as of version 1.0. Thanks to the \wilson{} project, the scope includes not only flavour physics observables, but in fact any observable where NP effects can be parametrized as dimension-6 Wilson coefficients in the SMEFT (or the weak effective  theory).
Observables that can play an important role in testing the SM and are currently not included in the package include
\begin{itemize}
\item Non-leptonic $B$ decays,
\item Rare $D$ decays,
\item atomic EDMs,
\item nuclear and neutron beta decays,
\item dijet and dilepton contact interaction searches at LHC,
\item four-lepton contact interaction searches at LEP,
\item flavour-blind CP-conserving low-energy precision measurements, e.g.\ atomic parity violation,
\item Higgs production and decay,
\end{itemize}
and many more.

Implementing more and more of these low-energy precision tests of the SM, \flavio{} can serve as the basis of a global
likelihood in EFT parameter space that can serve as a powerful tool to test extensions of the SM in the future \cite{smelli}.

\section*{Acknowledgements}

\noindent
I would like to thank
Peter Stangl,  Jacky Kumar, Christoph Niehoff, Ece G\"urler,
Matthew Kirk, Jason Aebischer,
Zeren Simon Wang, Stephanie Reichert,
Frederik Beaujean,
Matthias Sch\"offel,
and
Albert Puig
for code contributions,
Jason Aebischer,
Paula Alvarez Cartelle,
Wolfgang Altmannshofer, Frederik Beaujean, Christoph Bobeth, Joachim Brod,
Danny van Dyk, Martin Gorbahn,
James Gratrex,
Sebastian J\"ager,
Joel Jones-Perez,
Alexander Lenz, Stefan Meinel,
Satoshi Mishima, Ayan Paul,
Marie-H\'el\`ene Schune, Peter Stangl, Florian Staub,
Avelino Vicente,
Javier Virto,
and Roman Zwicky
for useful discussions and numerical cross-checks,
Miko\l aj Misiak for providing numerical results necessary for inclusive radiative
$B$ decays,
Paolo Gambino for providing numerical results necessary for inclusive semi-leptonic
$B$ decays,
and
Matthias Steinhauser and Ken Mimasu
for kind permission to temporarily reuse code from \texttt{CRunDec}
\cite{Chetyrkin:2000yt,Schmidt:2012az} and \texttt{Rosetta}
\cite{Falkowski:2015wza},
respectively, in older versions of \flavio{}.
Last but not least, I am indebted to the fantastic open source community
of the Python ecosystem for the awesome software this
project is built on, in particular the
numpy \cite{numpy},
scipy \cite{scipy},
matplotlib \cite{matplotlib},
IPython \cite{ipython},
and
Jupyter \cite{jupyter}
projects.
My work was supported by the DFG cluster of excellence ``Origin and Structure of the Universe''.

\bibliographystyle{JHEP}
\bibliography{bibliography}

\end{document}